# ProvLet: A Provenance Management Service for Long Tail Microscopy Data

Hessam Moeini, Todd Nicholson, Klara Nahrstedt, Gianni Pezzarossi
Coordinated Science Laboratory, University of Illinois at Urbana-Champaign, Urbana, IL 61801
Email: {moeini, tcnichol, klara, gpezza2}@illinois.edu

## ABSTRACT

Provenance management must be present to enhance the overall security and reliability of *long tail microscopy (LTM)* data management systems. However, there are challenges in provenance for domains with LTM data. The provenance data need to be collected more frequently, which increases system overheads (in terms of computation and storage) and results in scalability issues. Moreover, in most of scientific application domains a provenance solution must consider network-related events as well. Therefore, provenance data in LTM data management systems are highly diverse and must be organized and processed carefully.

In this paper, we introduce a novel provenance service, called ProvLet, to collect, distribute, analyze, and visualize provenance data in LTM data management systems. This means (1) we address how to filter and store the desired transactions on disk; (2) we consider a data organization model at higher-level data abstractions, suitable for step-by-step scientific experiments, such as datasets and collections, and develop provenance algorithms over these data abstractions, rather than solutions considering low-level abstractions such as files and folders. (3) We utilize ProvLet's log files and visualize provenance information for further forensics explorations. The validation of ProvLet with actual long tail microscopy data, collected over a period of six years, shows a provenance service which yields a low system overhead and enables scalability.

## 1 INTRODUCTION

Data collection is currently a big challenge in many scientific domains including materials science and semiconductor device fabrication. National Science and Technology Council (NSTC) reported a 20-years gap from discovery of new materials to fabrication of next-generation devices [1]. In contrast to other scientific domains with homogenous, well-organized data in an offline or batch manner, data in materials science are *long tail* data, i.e., small and medium sized data sets collected during day-to-day experimentation at microscopes such as SEM (Scanning Electron Microscope) or TEM (Transmission Electron Microscope). The current manual notetaking of complex experiments can lead to inconsistent or inadequate documentation. Figure 1 shows multiple text and image files that are necessary to capture all the pertinent data of a single experiment. Moreover, data transfer from tools (e.g., microscopes) is often done using flash-drive or emails that carry limitations and security risks. Because of these problems, scientists keep only the "best" results, and the "best" is determined by a narrow and specific scientific objective. The remaining data is often discarded which could contain information significant to others in the future.

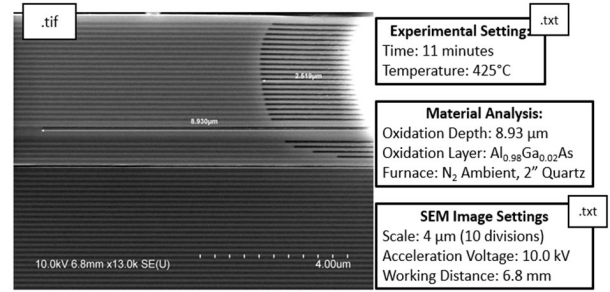

**Figure 1: Stored Metadata Describing Single Experiment**

The lack of a trusted data management system which provides real-time data capture, distribution, processing, managing, analyzing, and sharing of long tail scientific microscopy data causes significant delay in bridging the innovation across scientific disciplines. There are multiple data management systems designed and developed for application domains with organized batch data [2-4]. On the other hand, Clowder [5] is a web-based data management system designed to manage long tail data. It follows a data organization model which represents step-by-step scientific experiments and device fabrication processes. Its data organization model introduces a hierarchy of nested *collections*, *datasets*, and *files*. Files are equivalent to the file systems and represent experimental results data. A dataset is a grouping of files that have metadata capturing the preparation information of the experimental sample. A collection is a user defined group of datasets to organize the experiments and the nested structure of collections makes it possible to organize data the way users prefer. Clowder also supports horizontal grouping by introducing a *space* concept. A space, as shown in Figure 2, is a group of collections, datasets, and files with defined user access rights which can be used to define different projects.

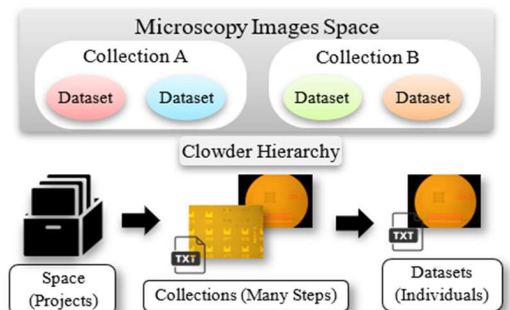

**Figure 2: Clowder: Robust LTM Data Organization**

Provenance (a.k.a. lineage) is a solution to improve system security and reliability in scientific lab environments. Data provenance records the history of data generation, data distribution, and its processing steps. There are different interesting applications for data provenance in long tail datasets such as (a) assessment of quality of data based on history of data in terms of its source and transformations, (b) audit trial of data for resource usage determination or detection of errors in data generation, (c) replication recipes, where a detailed provenance information can help scientists to repeat data derivation, (d) attribution, where scientist can establish ownership of data and determine liability in case of erroneous data, and (e) informational purpose of provenance data for discovery of data and interpretation of data.

There are different challenges for provenance management in LTM data management systems. Because of the nature of LTM data, the processes of collecting and distributing provenance data, processing them, storing the processed log files, securing the log files, retrieving the logs from the database, and analyzing them, are different from systems handling offline batch files. First, LTM data are captured and stored frequently in small bursts. Therefore, it is critical for a provenance solution to monitor the scientific experiment-running system continuously and log transactions (events) frequently. This expands the volume of provenance log files and increases system overheads. Second, provenance data size overheads and system scalability need to be addressed carefully. Third, often it is not sufficient to monitor and log transactions in the data management system. Provenance solution needs to address network-related transactions. Forth, transactions in LTM data management systems are of different types and provenance data are highly diverse. How to organize and store the provenance data effectively to query them efficiently is another challenging task.

Provenance management has been widely used in relation to art industry [6] as well as different scientific domains such as geology [7], bioinformatics [8], astronomy [9] computer networks [10], machine learning [11], supply chains [12], and et cetera. Most of these provenance management models are based on batch data considerations and are not able to address above challenges. Since the nature of data is different in these provenance solutions, they usually are offline solutions and only consider the lower level of data abstraction, the file system, and folders. Our goal is that a comprehensive provenance solution should consider not only observations of data/metadata files, but also monitoring of higher levels of data abstractions such as collections, datasets, and spaces.

This paper introduces a new service for provenance management in long tail microscopy (LTM) data, called ProvLet. The solution not only provides provenance-based security but also secures the provenance data itself. It ensures that all provenance data and processes are securely accessed, tracked, and archived on disks. Our contributions in this paper are:

- We present ProvLet, a provenance management solution for long tail microscopy data in materials science and semiconductor device fabrication research labs. We log any access of microscopy scientists to datasets (besides files), collections and spaces (not folders) which are the "data" driven abstractions how Clowder organizes scientific data. This is the new view on LTM-based provenance in comparison to provenance over files and folders of data.
- We implement ProvLet in an efficient manner. We collect diverse provenance data and network-related transactions to cover all LTM data manipulations in scientific labs. We address scalability issue and let system and admin users modify the frequency of recordings and customize the logs granularity to lower storage overhead. We also introduce a fast retrieval mechanism to read the provenance log files and search through the stored provenance data efficiently. Finally, we analyze and visualize the provenance data, and make it easier for scientists to query their log files and track the processes efficiently.
- We validate our provenance management solution by executing ProvLet on an actual microscopy dataset of LTM data collected from a scientific lab and over a period of six years.

Next, we introduce design and algorithmic details of ProvLet in Section 2. We then discuss implementation and validation details of ProvLet using real microscopy data in Section 3. We conclude in Section 4.

## 2 PROVLET PROVENANCE MANAGEMENT

In this Section, we first introduce ProvLet service architecture. Then, we discuss characteristics of ProvLet in order to address challenges discussed in Section 1.

### 2.1 ProvLet Service Architecture

Clowder is a data management system suitable for LTM data. It has a lightweight web uploader tool used to collect and upload scientific data and metadata to store them on the remote repository. In this work, we assume Clowder as the LTM data management system as well as the network packet analyzer and monitoring tools to assist ProvLet. Figure 3 shows the ProvLet architecture in details.

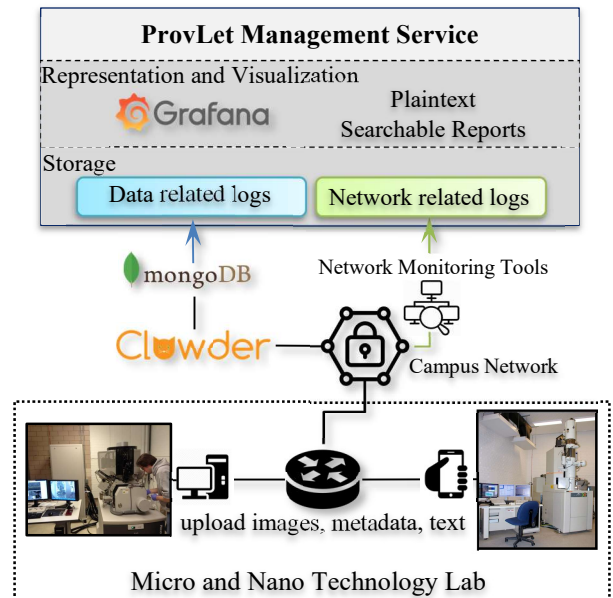

Figure 3: ProvLet: Provenance Service Architecture

Scientists in micro-and-nano-technology lab upload their microscopy data, and their corresponding metadata, using the Clowder's web-based uploader, through the campus network. ProvLet collects two different streams of provenance records, one coming from the Clowder data management system, and the other from network monitoring tools. ProvLet data coming from these two directions get filtered based on a desired granularity defined by admin users and aggregated to be stored on disks for further applications. These stored ProvLet data then can be queried and represented as a searchable plaintext report or in dashboard with visualized graphs and diagrams.

### 2.2 ProvLet Service Characteristics

In this Subsection, we discuss characteristics of ProvLet that address challenges for LTM data provenance management.

**Data Collection and Storage.** Clowder is a MongoDB-based system. All data and files in Clowder are stored using MongoDB which is a NoSQL database program. MongoDB uses JSON-like documents with optional schemas. In this work, we also use MongoDB to store and access ProvLet log files. Each transaction in the ProvLet log file has a *unique identifier*, a *timestamp*, *user information*, *object information* (if any) to specify a file, *dataset, collection, space*, or user, and finally an *event type*. Different types of events may include more detailed optional attributes. ProvLet examples of event types are shown in Figure 5(b).

**Scalability.** In order to address scalability and to reduce the provenance overhead (in terms of storage and the computing power required to process the log files), admin users can adjust the granularity of ProvLet log files and activate only a limited subset of all the available event types. This action adjusts the frequency of recording in the ProvLet log files and saves space for more important transactions. The pseudocode and explanation of ProvLet data storage algorithm is given in the following paragraph. This algorithm helps ProvLet to adaptively filter the collected records and address scalability issue by reducing system overheads.

Function "data-storage" receives a list of requested events, *req-events*, as the input. *req-events* is a list of different event types that admin user asked to be collected by ProvLet. Each event type in *req-events* has a priority value of *hpr*, *mpr*, or *lpr*, representing a high, medium, or low importance for the event type. ProvLet listens to both Clowder and network transactions to collect the information of a new event, *new-event*, when it occurs. It checks *new-event*'s event type to see if it has been requested by the admin users. If $new\text{-}event.type \in req\text{-}events$, ProvLet further checks the provenance file, *proveData*, and gets its current size. If $size(proveData)$ is lower than a defined Provenance Data Bounding (*PDB*) value, ProvLet appends the *new-event* to *proveData*. However, if adding the new event makes the size of provenance data higher than the *PDB* value, ProvLet takes actions to free-up the space as follows: It first generates a system alert to let admin users review the *req-events* list and revise it manually if needed. Also, another function "review-events()" checks the collected events and automatically revise *req-events* based on long tail microscopy data characteristics in a new event request list as *new-re*. In LTM data, frequency of capturing an event type can impact the priority of that event type. For example, as Figure 5 (b) illustrates, event type "update_dataset_information" has been recorded with a very high frequency. Then, ProvLet assigns a high priority to this event type since it indicates importance that a lot of changes are happening in the experiment over time. Other event types that are less frequent could have lower priorities. Finally, function "lowest-ranked-records" will be called to list all the lowest ranked current records in *proveData*, the *lrr*. This function ranks the collected events based on their *timestamp* and revised *priority* values. Recent high priority events have the highest rankings while older low priority data can be archived or ignored if necessary. When a data is required to be replaced, ProvLet first tries to archive it in a second repository, if available. Otherwise, it simply removes the lowest ranked records and stores more valuable information on disk. Function "data-storage" returns *provData* for further applicairons and queries.

---

**Algorithm 1:** ProvLet Data Storage

data-storage(*req-events*)
1  $provData = \varnothing$;
2  monitor network and Clowder to capture *new-event*;
3  **while** ($new\text{-}event.type \in req\text{-}events$) **do**
4   **if** ($size(proveData) < PDB$) **then**
5    append *new-event* to *provData*;
6   **else**
7    generate-alert();
8    *new-re* = review-events(*req-events*);
9    *lrr* = lowest-ranked-records(*provData*, *new-re*);
10    **if** ($new\text{-}event.rnk > lrr.rnk$) **then**
11     move *lrr* to a second repository;
12     go to step 5;
13    **else**
14     ignore *new-event*;
15  **return** (*provData*);

---

**Network Provenance.** Some of the informative data in ProvLet is related to the networks connecting users to Clowder. For example, it is critical to know the address of a machine being used to upload a file. We monitor the campus network where users connect to work with Clowder and log the desired information for further references. An admin user can process these log files and for example understand the machine information (MAC and IP address) that a scientist used to login to the system and to do her/his desired actions. The attributes we consider for network provenance are *timestamp, source IP address, packet destination, used network protocol, length of packet*, and any other additional information (application layer info, fragmentation info, etc.) if available.

**Securing Provenance Data.** We secure the ProvLet data via access control mechanisms. Only a permitted user (lab admin, IT admin, research funding principal investigator, etc.) can access ProvLet log files and process these data to get forensics information needed for audits, backtracking the current state, troubleshooting of a problem, checking the ownership of a specific outcome and result, measuring the user utilization for accountability, and other usages as discussed in Section 1. A regular user has only access to ProvLet data in relation with her/his own transactions.

**Representation Model and Data Visualization.** Users can use the stored ProvLet data and achieve their desired goals. Plaintext JSON-like formatted log files can be parsed and used for advanced searches and queries. However, the number of transaction records in the log file can grow exponentially (by adding new users to the system or increasing the granularity of log files). Moreover, analyzing, and querying databases and log files is not always an easy task for users with other specialties. We visualize the provenance data in a pre-designed dashboard and give different views of this dashboard to different users based on their granted access level. Admin users can easily customize these charts and reports in the dashboard based on their requirements and preferences.

## 3 PROVENANCE GRAPH

## 4 EXPERIMENTAL STUDY

We validate the ProvLet model by deploying ProvLet service within the 4CeeD data platform [13]. 4CeeD is an enhanced Clowder data management system for materials scientists to capture, curate, coordinate, correlate, and distribute laboratory data. Alongside with the lightweight web uploader tool used for data collection, 4CeeD has a web-based curator to organize and manage data files and metadata. We used this platform and collected, filtered, and customized data provenance log files in an active materials science laboratory over a period of six years (2015-2020). We have also used Wireshark [14] and tcpdump [15] to monitor and log network packets. ProvLet visualizes log files using different reports and diagrams designed in dashboards on Grafana [16], an open-source analytical platform which let us query, visualize, alert on, and understand different metrics.

The 4CeeD instance includes LTM data with 150 spaces, 730 collections, 2425 datasets and 24259 files created and uploaded by 192 different users. Total volume of data and metadata stored on this instance is equal to 297 GB. We have collected 5.7 MB of provenance data over this period of six years. Clearly, the required space size for provenance log files is much smaller than the actual LTM data, so the overall provenance overhead is very low ($\approx \% 0.002$). This is a fact that the size of provenance log files is relatively small compared to the disk volume required by LTM data management system; however, it is very crucial to efficiently collect events and lower the computational overheads in provenance management service as discussed earlier.

We first consider separate spaces as different projects and explore the time needed to complete projects with LTM data. We calculate project duration based on the date spaces are created and the date they and their subsequent collections, datasets, and files are lastly updated. Figure 4 provides the highest, lowest and the average required calendar days for projects in each calendar year and based on the projects' starting dates. Notice that projects in later years 2019 and 2020, may be still active and the actual duration time can change in future.

Figure 4 emphasizes that a scientific experiment with LTM data usually lives much longer periods than just few minutes. It can last months to years and therefore data support the long duration of experiments demanding long duration of data management.

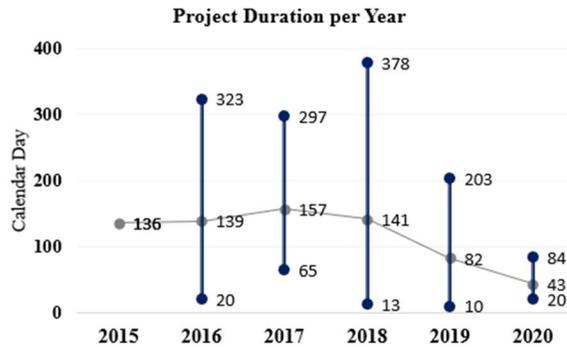

Figure 4: Highest, Lowest, and The Average Project Duration Times for Different Years

We then study the number of transactions and active users per each year. As can be seen in Figure 5 (a), both metrics captured the highest activities in 2018. This information can help the lab managers and IT admins to understand the system performance and plan accordingly. For example, one may need to explore the reasons of the decrease in number of active users in year 2019. We finally explore the occurrence quantity of different event types and look at the frequency of different events in real-life scenarios. Figure 5 (b) ranks the most frequent transactions from 33 different event types defined by ProvLet. This ranking very much represents the frequency of small files' accesses, frequency of the accesses to data, and updating data over long period of time as experiments last over months and even years. It will be used in "review-events" function explained in Subsection 2.2 to rank and replace the LTM data provenance records on disk and free-up some spaces when needed.

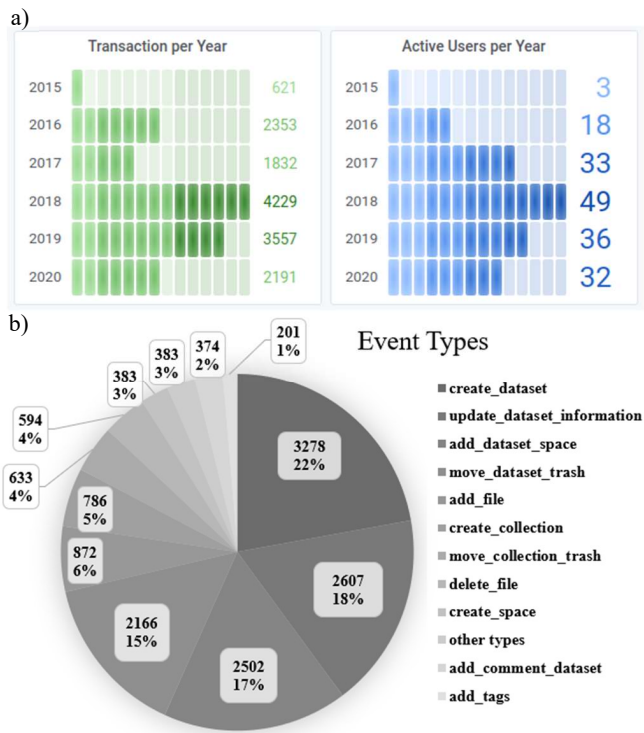

Figure 5: (a) Events and Active Users Numbers per Year
(b) Different Event Types Occurrence in Log Files

## 5 CONCLUSIONS

In this work we discussed existing challenges on managing provenance for heterogeneous LTM data management systems as they are very useful in managing data and metadata in specific scientific domains like materials science and semiconductor device fabrication. We introduced ProvLet, a provenance solution, to address these challenges and discussed its architecture in detail. We have collected data and network related provenance data adaptively and stored them in ProvLet log files for further explorations. We have validated ProvLet with actual provenance data collected for a period of six years in an active materials science lab.

## ACKNOWLEDGEMENT

This work is supported by the National Science Foundation under Grant No. ACI-1835834.

## REFERENCES

[1] John P. Holdren, 2011. Materials genome initiative for global competitiveness. *National Science and Technology Council OSTP*. Washington, USA.

[2] Bertram Ludäscher, Ilkay Altintas, Chad Berkley, Dan Higgins, Efrat Jaeger, Matthew Jones, Edward A. Lee, Jing Tao, and Yang Zhao. 2006. Scientific workflow management and the Kepler system. *Concurrency and computation: Practice and experience* 18, no. 10 (August 2006), 1039–1065.

[3] Ewa, Deelman, Karan Vahi, Mats Rynge, Rajiv Mayani, Rafael Ferreira da Silva, George Papadimitriou, and Miron Livny. 2019. The evolution of the pegasus workflow management software. *Computing in Science & Engineering* 21, no. 4 (2019), 22–36. DOI: https://doi.org/10.1109/MCSE.2019.2919690

[4] Tom Oinn, Mark Greenwood, Matthew Addis, M. Nedim Alpdemir, Justin Ferris, Kevin Glover, Carole Goble, Antoon Goderis, Duncan Hull, Darren Marvin, Peter Li, Phillip Lord, Matthew R. Pocock, Martin Senger, Robert Stevens, Anil Wipat, and Chris Wroe. 2006. Taverna: lessons in creating a workflow environment for the life sciences. *Concurrency and computation: Practice and experience* 18, no. 10 (2006), 1067–1100. DOI: https://doi.org/10.1002/cpe.993

[5] Luigi Marini, Indira Gutierrez-Polo, Rob Kooper, Sandeep Puthanveetil Satheesan, Maxwell Burnette, Jong Lee, Todd Nicholson, Yan Zhao, and Kenton McHenry. 2018. Clowder: Open Source Data Management for Long Tail Data. In *Proceedings of the Practice and Experience on Advanced Research Computing (PEARC '18)*. Association for Computing Machinery, New York, NY, USA, Article 40, 1–8. DOI: https://doi.org/10.1145/3219104.3219159

[6] Qiuyan Lu, Guisen Zou, Yanxiang Li, Lin Zheng, and Wei Wang. 2020. Provenance study on 'Big bronze drums': a method to investigate the ancient bronze industry of Guangxi, Southwest China from Han to Tang dynasty (around 200 BC–900 AC). *Journal of Cultural Heritage* (2020). DOI: https://doi.org/10.1016/j.culher.2020.02.002

[7] James Frew, and Rajendra Bose. 2001. Earth system science workbench: A data management infrastructure for earth science products. In *Proceedings Thirteenth International Conference on Scientific and Statistical Database Management*. IEEE SSDBM 2001, pp. 180–189. DOI: https://doi.org/10.1109/SSDM.2001.938550

[8] Polyane Wercelens, Waldeyr da Silva, Klayton Castro, Aleteia PF Araujo, Sergio Lifschitz, and Maristela Holanda. 2019. Data Provenance Management of Bioinformatics Workflows in Federated Clouds. In *2019 IEEE International Conference on Bioinformatics and Biomedicine (BIBM)*. IEEE. 750–754. DOI: https://doi.org/10. 1109/BIBM47256.2019.8983373

[9] Anastasia Galkin, Kristin Riebe, Ole Streicher, Francois Bonnarel, Mireille Louys, Michèle Sanguillon, Mathieu Servillat, and Markus Nullmeier. 2018. Provenance for astrophysical data. In *International Provenance and Annotation Workshop*. Springer, Cham. 252–256. DOI: https://doi.org/10.1007/978-3-319-98379-0_30

[10] Wenchao Zhou, Micah Sherr, Tao, Xiaozhou Li, Boon Thau Loo, and Yun Mao. 2010. Efficient querying and maintenance of network provenance at internet-scale. In *Proceedings of the 2010 ACM SIGMOD International Conference on Management of data (SIGMOD '10)*. Association for Computing Machinery, New York, NY, USA, 615–626. DOI: https://doi.org/10.1145/1807167.1807234

[11] Yinjun Wu, Val Tannen, and Susan B. Davidson. 2020. PrIU: A Provenance-Based Approach for Incrementally Updating Regression Models. In *Proceedings of the 2020 ACM SIGMOD International Conference on Management of Data (SIGMOD '20)*. Association for Computing Machinery, New York, NY, USA, 447–462. DOI: https://doi.org/10.1145/3318464.3380571

[12] Henry M. Kim, and Marek Laskowski. 2018. Toward an ontology-driven blockchain design for supply-chain provenance. *Intelligent Systems in Accounting, Finance and Management* 25, no. 1 (2018): 18-27. DOI: https://doi.org/10.1002/isaf.1424

[13] Phuong Nguyen, Steven Konstanty, Todd Nicholson, Thomas O'Brien, Aaron Schwartz-Duval, Timothy Spila, Klara Nahrstedt, Roy H. Campbell, Indranil Gupta, Michael Chan, Kenton Mchenry, and Normand Paquin. 2017. 4CeeD: Real-Time Data Acquisition and Analysis Framework for Material-Related Cyber-Physical Environments. In *2017 17th IEEE/ACM International Symposium on Cluster, Cloud and Grid Computing (CCGRID)*. 11–20. DOI: https://doi.org/10.1109/ CCGRID.2017.51

[14] Wireshark. Available: https://www.wireshark.org, accessed September 2020.

[15] Tcpdump. Available: https://www.tcpdump.org, accessed September 2020.

[16] Grafana. Available: https://grafana.com, accessed September 2020.